\documentclass[12pt]{article}
\usepackage{paper2e}
\usepackage{mydefs2e}
\usepackage{epsf}

\begin{document}
\begin{titlepage}

\renewcommand{\thefootnote}{\fnsymbol{footnote}}

\preprint{LBNL-49280,\ \ UM-PP-02-024}

\begin{center}
\Large\bf
Realistic Anomaly Mediation\medskip\\
with Bulk Gauge Fields
\end{center}

\author{Z. Chacko}
\address{Department of Physics, University of California\\
Berkeley, California 94720}
\vskip -0.1in
\centerline{and}
\vskip .15in
\address{
Theory Group, Lawrence Berkeley National Laboratory\\
Berkeley, California 94720}

\author{Markus A. Luty}
\address{Department of Physics, University of Maryland\\
College Park, Maryland 20742}

\begin{abstract}

We present a simple general framework for realistic models of
supersymmetry breaking driven by anomaly mediation. We consider a
5-dimensional `brane universe' where the visible and hidden sectors are
localized on different branes, and the standard model gauge bosons
propagate in the bulk. In this framework there can be charged scalar
messengers that have contact interactions with the hidden sector, either
localized in the hidden sector or in the bulk. These scalars obtain soft
masses that feed into visible sector scalar masses at two loop order via
bulk gauge interactions. This contribution is automatically flavor-blind,
and can be naturally positive.  If the messengers are in the bulk this
contribution is automatically the same order of magnitude
as the anomaly mediated
contribution, independent of the brane spacing. If the messengers are
localized to a brane the two effects are of the same order for relatively
small brane spacings.  The gaugino masses and $A$ terms are determined
completely by anomaly mediation. In order for anomaly mediation to
dominate over radion mediation the radion must be is stabilized in a
manner that preserves supersymmetry, with supergravity effects included.
We show that this occurs in simple models. We also show that the $\mu$
problem can be solved by the vacuum expectation value of a singlet in this
framework.

\end{abstract}


\end{titlepage}

\renewcommand{\thefootnote}{\arabic{footnote}}

\section{Introduction}
Anomaly mediated supersymmetry breaking (AMSB) \cite{RS0} is an attractive
framework for solving the supersymmetric flavor problem.
AMSB requires only that non-gravi\-tational interactions (including
Planck-suppressed contact interactions between the visible and
hidden sectors) are negligible compared to the interactions of
low-energy supergravity.
This can naturally occur in models with extra dimensions in which the visible
and hidden sectors are localized on different branes \cite{RS0}, or in models
where a conformal sector dynamically suppresses the contact terms
\cite{compdim}.
In AMSB, the leading contribution to supersymmetry (SUSY) breaking
parameters is a model-independent supergravity effect closely related to the
conformal anomaly \cite{RS0,GLMR}.
The SUSY breaking masses are automatically flavor-blind, and are of order
$m_{3/2} / 16\pi^2$.

If the visible sector is the minimal supersymmetric standard model
(MSSM), and there are no interactions between the visible and hidden sectors
other than supergravity, then the slepton mass-squared terms are negative
\cite{RS0}.
In this paper we discuss a very simple and general framework that can
naturally explain why the slepton masses are positive and of the correct
magnitude, while preserving the gaugino mass predictions of minimal AMSB.
Our proposal is closely related to that of \Ref{GAMSB}, but is more general
and robust, as we will explain.
For other proposals for realistic models
based on AMSB, see \Refs{realAMSB}. For potential experimental signals of
AMSB, see for example \Refs{experiment}.


The idea is very simple.
Consider a 5D model in which the visible sector gauge bosons propagate in the
bulk.
In this scenario contact terms between the gauge fields and the hidden sector
are not suppressed by higher-dimensional locality, but they are naturally
suppressed if there are no singlets in the hidden sector.
We also assume that the extra dimension is stabilized in a manner that does not
break SUSY.
(We will show that this occurs in simple models.)
The leading contribution to the gaugino masses and $A$ terms
then comes from anomaly mediation.
This framework for anomaly mediation was previously discussed in \Ref{GAMSB},
where it was shown that contact terms between the bulk gauge fields and the
hidden brane can generate visible scalar masses of the same order as
anomaly mediation.

%

In this paper we point out an additional contribution to the visible
scalar masses in this framework.
Because the standard model gauge bosons propagate in the bulk, there can be
additional charged scalars, either in the bulk or on the hidden brane, that
have contact terms with fields localized on the hidden brane.
One particularly natural candidate is the scalar component of the bulk
gauge supermultiplet.
Up to volume factors, these scalars will get a mass of order $m_{3/2}$
from contact interactions with the SUSY breaking sector.%
\footnote{We assume that the bulk spacetime is nearly flat,
\ie the warp factor is not significant.}
These in turn contribute to visible scalar masses at 2 loops.
The contribution from these scalar messengers
comes from gauge loops, and is therefore automatically
flavor-blind.
The sign of the visible mass-squared depends on the coefficient of a
higher-dimension operator on the hidden brane, and can naturally be
positive.
For scalar messengers localized on the hidden brane, the messenger
contribution to visible scalar masses is the same size as the AMSB
contribution only for a special value of the compactification scale,
as in \Ref{GAMSB}.
However, if the messengers propagate in the bulk, the messenger contribution
to visible scalar masses is naturally the same as the AMSB contribution.
This is independent of the compactification scale, and independent of whether
the fundamental theory is strongly or weakly coupled.
We therefore regard the framework of bulk scalar messengers as a very
attractive and robust framework for realistic models of AMSB.



We also show that in this framework the $\mu$ problem can be naturally
solved in the context of the `next-to-minimal supersymmetric standard
model' (NMSSM).
This is a model with a singlet at the weak scale whose VEV gives rise
to an effective $\mu$ term.
The singlet mass-squared can be driven negative at the weak scale by
radiative corrections, so all weak scale VEV's are explained by radiative
symmetry breaking.


This paper is organized as follows.
In section 2, we introduce models of scalar messengers and estimate
the SUSY breaking parameters.
In section 3, we address the $\mu$ problem.
In section 4, we discuss radion stabilization.
Section 5 contains our conclusions.

\section{Scalar Messengers}
As discussed above, we assume that the standard-model gauge multiplets
propagate in the bulk.
A 5D gauge multiplet consists of the gauge field $A_M$ ($M = 0, \ldots, 4$),
a real adjoint scalar $\si$, and a fermion $\la$.
We assume that the $5^{\rm th}$ dimension is compactified on a $S^1 / Z_2$
orbifold of radius $r$.
The fixed points of the orbifold are `branes' on which the
hidden and visible sectors can be localized.
The $Z_2$ parity assignments of the gauge field assigned
so that $A_5$, $\si$, and half of the $\la$ components are odd.
These states will then get masses of order $1/r$, and the
surviving degrees of freedom make up an $\scr{N} = 1$ gauge multiplet.
(See \eg \Refs{MP,AGW} for details.)

In order to proceed, we must estimate higher-dimension contact terms that come
from new physics at the fundamental scale.
In the spirit of string unification, we will assume that there is a
fundamental scale $\La$ such that all quantum corrections (whether from
brane or bulk loops) are of order $\ep$.
The value $\ep \sim 1$ corresponds to strong coupling at the fundamental
scale, while $\ep \sim 1/\ell_4 = 1/16\pi^2$ corresponds to the coupling
strength of a 4D theory with all couplings of order unity.%
\footnote{The counting used \eg in \Ref{GAMSB} assumes that all
couplings are order 1 in units of the 5D Planck scale $M_5$,
and that heavy bulk states have mass of order $M_5$.
However the 5D gauge couplings
must be larger than this estimate by a factor of 10 to account for
the fact that $g_4 \sim 1$.}
The power counting of factors of $\ep$ and loop factors is explained
in \Ref{CLPU1}.

The 4D gauge couplings are given by
\beq
g_4^2 \sim \frac{g_5^2}{r} \sim \frac{\ep \ell_5}{\La r},
\eeq
where $\ell_5 = 24\pi^3$ is the 5D loop factor.
   From the fact that $g_4^2 \sim 1$ we can fix the radius:
\beq
\La r \sim \ep \ell_5.
\eeq
Since $\La$ naturally sets the scale for heavy states in the bulk,
so we require $\La r \gsim 10$ for the absence of flavor-changing
neutral currents (FCNC's).
The acceptable range for $\ep$ is therefore
\beq
10^{-2} \lsim \ep \lsim 1.
\eeq
With this estimate, we find (somewhat amusingly)
\beq
\La \sim M_4,
\eeq
where $M_4 \simeq 2 \times 10^{18}\GeV$ is the 4D Planck scale.

We find the strong coupling scenario $\ep \sim 1$ particularly attractive,
since the small value of the observed gauge couplings is explained
by the same large radius that suppresses SUSY flavor violation.
Also, strong coupling at the fundamental scale may be required to solve
general difficulties with weakly-coupled string vacua \cite{DineSeiberg}.

%
%
%
%
%

\subsection{Hidden Scalar Messengers}
Since the visible gauge fields are in the bulk, we can have charged chiral
fields $\Phi$ localized on the hidden brane.
We now explore the possibility that these play the role of the hidden scalar
messengers.

The hidden fields $\Phi$ will
in general have couplings to the field $X$ whose $F$ component
breaks SUSY:
\beq[phiop]
\De\scr{L}_5 \sim \de(x^5 - x^5_{\rm hid}) \myint d^4\th\,
\frac{\ep \ell_4}{\La^3} X^\dagger X \Phi^\dagger \Phi.
\eeq
These give a scalar mass-squared term
\beq
m_{\tilde\Phi}^2 \sim \ep \ell_4 m_{3/2}^2.
\eeq
%
%
The sign of $m_{\tilde\Phi}^2$ is determined by the sign of the
operator \Eq{phiop},
which arises from unknown short-distance physics.
We also require the field $\Phi$ to have a supersymmetric mass
$M_\Phi$ to give mass to the fermion components of $\Phi$.
This supersymmetric mass can arise from a brane-localized mass term on the
hidden brane.

The effect of the supersymmetry breaking messenger mass on the visible
sector scalar mass via 2-loop diagrams.
This effect is analogous to $D$-type gauge mediation in four dimensions
\cite{PT}.
Since it arises from gauge interactions it is flavor blind.
The size of the visible scalar mass depends on the relative magnitude
of $M_\Phi$
and the compactification scale $1/r$.
Let us first consider the case $M_\Phi \gg 1/r$.
Then we can integrate out the field $\Phi$ in the 5D theory.
In the effective lagrangian below the scale $M_\Phi$ the leading terms
that depends on $m_{\tilde\Phi}^2$ have the form
\beq
\De\scr{L}_5 \sim \de(x^5 - x^5_{\rm hid}) \myint d^4\th\,
\frac{\ep \ell_5}{\ell_4}
\, \frac{1}{(M_\Phi / Z_\Phi)^2 \La}
\bar{W} \bar{D} D W,
\eeq
where $D$ is the SUSY covariant derivative and
$Z_\Phi = 1 - \th^4 m_{\tilde{\Phi}}^2$.
Expanding out the $m_{\tilde{\Phi}}^2$ term in $Z_\Phi$, we obtain
precisely the `gaugino assisted' operators of \Ref{GAMSB},
but here they are suppressed by $M_\Phi$ rather than the fundamental
scale.
This operator gives rise to a 1-loop contribution to visible scalar masses
of order
\beq
\De m_{\tilde{Q}, \tilde{L}}^2 \sim
\frac{\ep \ell_4}{(M_\Phi r)^2}\,
\frac{m_{3/2}^2}{\ell_4^2}.
\eeq
(See \Ref{GAMSB} for a detailed calculation.)
%
%
This can be the same order as the anomaly mediated
contribution for the right value of $M_\Phi$.

We obtain a somewhat more robust result in the case $M_\Phi \ll 1/r$.
In this case the 4D effective theory below the compactification
scale includes $\Phi$.
The 2-loop running between the compactification scale and the mass $M_\Phi$
gives a log-enhanced contribution to the visible scalar masses
\beq
\De m_{\tilde{Q}, \tilde{L}}^2 \sim
\ep \ell_4 \ln(M_\Phi r) \,
\frac{m_{3/2}^2}{\ell_4^2}.
\eeq
%
%
This naturally gives a flavor-blind positive contribution to the squark and
slepton mass-squared terms that is somewhat larger than the AMSB contribution.
(Remember $\ep \ell_5 \sim \La r \gsim 10$.)
To get a positive log-enhanced contribution to the visible scalar mass, we
require $m_{\tilde\Phi}^2 < 0$, so we require $|M_\Phi| >
|m_{\tilde{\Phi}}|$ in order for the potential to be stable at $\Phi = 0$.
In this case, the messenger contribution is the same size as the AMSB
contribution for $\ep \sim 1/\ell_4$, which is what we obtain if we
extrapolate the weak coupling of the standard model to the fundamental scale.
The results are only logarithmically sensitive to the SUSY mass $M_\Phi$.

\subsection{Bulk Scalar Messengers}
The mechanism discussed above is very simple and general, but it requires
the introduction of a new supersymmetric mass scale $M_\Phi$.
Also, the new contribution to the visible scalar masses is larger than the
AMSB contribution, especially for the attractive case where the fundamental
theory is strongly coupled.
Both of these potential difficulties can be elegantly removed if the
scalar messengers propagate in the bulk.
We then obtain a very robust framework for realistic AMSB that is the
central result of this paper.

We can naturally obtain a SUSY mass for a bulk scalar of order
the compactification scale $1/r$, in at least two ways.
One way is to impose enough orbifold boundary conditions such that all
bulk
scalars
are odd under at least one orbifold projection.
Another way is to add a Planck-scale mass term localized on one of the branes;
the lightest scalar KK mode is then `repelled' from the brane with the
mass term and gets a mass of order $1/r$ \cite{CLPU1},\cite{AHNSW}.
With either mechanism, bulk scalar messengers do not require the introduction
of an additional supersymmetric mass scale.

Let us consider these mechanisms in more detail for the the
case of a bulk hypermultiplet.
%
Using the formalism of \Ref{AGW} this can be parameterized by two
independent $\scr{N} = 1$ chiral superfields $\Phi$ and $\bar{\Phi}$ that
depend on $x^5$.

Let us first consider an orbifold symmetry of the type
$x_0^5 +  x^5 \mapsto x_0^5 - x^5$ to give mass to the bulk
hypermultiplets.
The kinetic term is given by
\beq
\scr{L}_5 = \myint d^4\th\,
(\Phi^\dagger \Phi + \bar{\Phi}^\dagger \bar{\Phi})
+ \left( \myint d^2\th\, \bar\Phi \partial_5 \Phi + \hc \right).
\eeq
The second term implies that if $\Phi$ is odd under such a parity, then
$\bar{\Phi}$ must be even, and {\it vice versa\/}.
We can therefore compactify on a $S^1 / (Z_2 \times Z_2)$ orbifolds, where
the two $Z_2$ reflections leave invariant $x^5_{\rm vis}$ and
$x^5_{\rm hid}$, respectively.
We choose the fields to have the following parity assignments:
$\Phi \sim (-, +)$, $\bar{\Phi} \sim (+, -)$.
This all components of the hypermultiplet a SUSY mass of order $1/r$.

Another way to give a SUSY mass to the even hypermultiplet when we have
only one $Z_2$ symmetry is to add
a brane-localized `mass' term of the form
\beq
\De\scr{L} \sim \de(x^5 - x^5_{\rm vis}) \myint d^2\th\,
\frac{\ell_5}{\ell_4} \La \tr(\Phi^2).
\eeq
This is allowed by gauge symmetry \eg\ if $\Phi$ is an
adjoint.
The effect of this term is to `repel' the wavefunction of $\Phi$ and
away from the visible brane by a factor of $\ell_4 /
(\ell_5 \La r)$
and give the lowest-lying KK modes of $\Phi$ a mass of order $1/r$
\cite{CLPU1}.
\footnote{This result can be understood from an elementary
electrostatic argument \Ref{CLPU1}.}

We assume that $\Phi$ has couplings to the hidden
brane of the form
\beq[Phihid]
\De\scr{L} \sim \de(x^5 - x^5_{\rm hid}) \myint d^4\th\,
\frac{\ep\ell_5}{\La^3}\, X^\dagger X \Phi^\dagger \Phi.
\eeq
(This coupling is allowed in the orbifold scenario, and is unsuppressed in
the scenario with large visible `mass' term.)
Remarkably, the 2-loop D-type gauge mediation diagrams with one insertion
of the operator above give rise
to visible scalar masses of order
\beq
m_{\tilde{Q}, \tilde{L}}^2 \sim \frac{m_{3/2}^2}{\ell_4^2},
\eeq
the same as the AMSB contribution.
Note that this is independent of $\ep$, which tells us how strongly coupled
the fundamental theory is.
Since $\ep\ell_5 \sim \La r$, this means the result is also independent of
the compactification radius.
The result is therefore very robust.

There are also are couplings of the bulk messengers to the
visible fields that violate flavor.
We now show that the flavor changing neutral currents induced by
these operators are consistent with experimental constraints.

In the orbifold model, the coupling of the $\Phi$ fields to the visible
sector is suppressed because $\Phi$ is odd under the $Z_2$ reflection
about the visible brane.
The leading direct coupling to visible fields
has the form
\beq[Phivisorb]
\De\scr{L} \sim \de(x^5 - x^5_{\rm vis}) \myint d^4\th\,
\frac{\ep \ell_5}{\La^5}\,
Q^\dagger Q
\partial_5 \Phi^\dagger \partial_5 \Phi.
\eeq
This gives rise to flavor-changing scalar masses from a 1-loop
diagram connecting this coupling to the coupling \Eq{Phihid}.
The result is
\beq
\De m^2_{\tilde{Q}} \sim \frac{\ell_4}{(\ep \ell_5)^4}\,
\frac{m_{3/2}^2}{\ell_4^2}.
\eeq
This gives 
unobservably small FCNC's
for strong coupling, and can give FCNC's near the experimental couplings
for weak coupling.
(Recall that $\ep \ell_5 \sim \La r \gsim 10$.)
There is a contribution of the same size coming from the $\bar{\Phi}$ field,
which has orbifold suppressed couplings to the hidden brane, but unsuppressed
couplings to the visible brane.

In the `large brane mass' scenario, the 1-loop contribution to the
flavor-changing
mass of squarks is suppressed by the KK wavefunction factor.
The resulting mass is of order
\beq
\De m^2_{\tilde{Q}} \sim \frac{\ell_4^3}{\ep^4 \ell_5^6}\,
\frac{m_{3/2}^2}{\ell_4^2},
\eeq
which gives 
unobservably small FCNC's for
for strong coupling, and can give FCNC's near the experimental couplings
for weak coupling.

%

Another very natural candidate for the bulk messenger is the adjoint scalar of
the bulk gauge multiplet.
Recall that the bulk gauge multiplet contains 2 4D scalars
$A_5$ and $\si$.
The field $A_5$ cannot have non-derivative couplings by 5D gauge
invariance, but non-derivative couplings of $\si$ are allowed.
To analyze the couplings of the bulk gauge multiplet to the branes,
we use the formalism of \Ref{AGW}.
The bulk gauge multiplet is parameterized by $\scr{N} = 1$ superfields $V$,
$\Si$ that depend explicitly on $x^5$.
Here $V$ is a vector superfield and $\Si$ is an adjoint chiral superfield with
\beq
V = \cdots + \bar{\th} \si^\mu \th A_\mu + \cdots,
\qquad
\Si = \sfrac{1}{2} (\si + i A_5) + \cdots.
\eeq
Under holomorphic gauge transformations $g$ these transform as
\beq[gaugesym]
e^{V} \mapsto g^{-1 \dagger} e^V g^{-1},
\qquad
\Si \mapsto g \Si g^{-1} - g \partial_5 g^{-1}.
\eeq
We can then form the combination
\beq
\tilde{\Si} = \Si + e^{-V} \Si^\dagger e^V - e^{-V} \partial_5 e^V
= \si + \cdots,
\eeq
which transforms as a gauge adjoint:
\beq
\tilde{\Si} \mapsto g \tilde{\Si} g^{-1}.
\eeq
Note that $\tilde{\Si}$ is not holomorphic, and therefore cannot appear in
superpotential terms.

We can therefore write couplings to the hidden sector of the form
\beq[Siop]
\De\scr{L}_5 \sim \de(x^5 - x^5_{\rm hid})
\myint d^4\th\, \frac{\ep \ell_5}{\La^3}\,
X^\dagger X \tr(\tilde{\Si}^\dagger e^V \tilde{\Si} e^{-V}).
\eeq
The estimate of the visible scalar masses is the same as for the bulk
hypermultiplet case, so we again obtain scalar masses of the
same order as the AMSB contribution.
We must also worry about flavor-violating operators similar to \Eq{Phivisorb},
and the estimates for FCNC's are again the same.

The discussion above assumes that the bulk gauge fields must have a SUSY mass
of order $1/r$.
As above, this can be acheived either by an orbifold projection, or by
using a brane-localized `mass' term for $\Si$.
The estimates for FCNC's are the same as above.
If we impose an orbifold projection, then $A_5$ (and hence $\Si$) must be
odd since the covariant derivative $D_5 = \partial_5 + i A_5$ is odd.
In this case we require that the hidden sector be on a brane that is
located away
from orbifold fixed points.
This means that the position of the hidden brane is dynamical, and gives rise
to a modulus that must be stabilized.
We will not address this problem here, but it does not appear unnatural for
such a brane to be stabilized between the fixed points of a $S^1 / Z_2$
orbifold.

\section{The $\mu$ Problem}
We now consider the $\mu$ problem in the present framework.
In AMSB an explicit $\mu$ term gives $B \sim m_{3/2}$, which is much larger
than anomaly-mediated masses of order $m_{3/2} / 16\pi^2$.
Therefore, in AMSB models the $\mu$ problem cannot be circumvented by simply
adding a tree-level $\mu$ term.
Here we point out that a $\mu$ term of the correct size can naturally
be generated by the vacuum expectation value of a singlet $S$
on the visible brane, with superpotential couplings
\beq[NMSSMW]
\De\scr{L}_5 = \de(x^5 - x^5_{\rm vis}) \myint d^2\th \left[
\la S H_u H_d + \frac{k}{3} S^3 \right].
\eeq
%
Note that the AMSB contribution to $m_S^2$ comes from the
superpotential couplings $\la$ and $k$, and is therefore positive.
The hidden messengers do not contribute to the $S$ mass (at leading 2
loop order)
because $S$ is uncharged, so $m_S^2 > 0$ at the compactification scale.
However, the 4D RG evolution to the weak scale can make $m_S^2 < 0$ at
the weak scale.
Also the $A$ terms generated by AMSB favor a nonzero VEV for $S$.

We assume that the compactification scale is higher than the GUT scale, and
that GUT threshold corrections are negligible.
In this case the hidden messenger corrections consist of a universal scalar
mass-squared $\De m_0^2$ for all squarks and sleptons (assuming a
$SO(10)$ GUT) and a
correction $\De m_H^2$ for the electroweak Higgs scalars.
In simple models, the ratio $\De m_H^2 / \De m_0^2$ is a ratio of GUT
Casimirs.
However, in realistic GUT's the electroweak Higgs is generally a mixture
of GUT representations, and we treat the ratio $\De m_H^2 / \De m_0^2$
as a free parameter.
With these simplifying assumptions, the parameters of this simplified model are
therefore the couplings $\la$, $k$, the top Yukawa coupling $y_t$,
and the mass parameters $\De m^2_0$ and $\De m^2_H$, all renormalized
at the GUT scale, and the SUGRA order parameter $\avg{F_\phi} \sim m_{3/2}$
that sets the size of the AMSB SUSY breaking.
(We do not consider large $\tan\be$ solutions, so we can neglect $y_b$
and $y_\tau$.)
These parameters are constrained by the observed value of the $W$ and
$Z$ masses, and the top quark mass.

In this framework, it is not difficult to find solutions which
satisfy all experimental
constraints with fine tuning of order $1\%$.
(The largest fine-tuning is in $\al_3$.)
We have not performed a systematic analysis of the parameter space, but
we make a few comments on the solutions we have found.
The solutions we find have $\De m^2_H > \De m^2_0$, which appears to be
required in order to obtain $m_S^2 < 0$ at the weak scale
(radiative symmetry breaking).
This can be natural if the electroweak Higgs has an admixture of a large
GUT representation. 
The presence of the $S$ field means that this model has an additional
neutralino that can mix with the other neutralinos and spoil the
minimal AMSB prediction of wino LSP.
However, it is not difficult to find realistic solutions in which the LSP
is mostly wino, so this is still a potential signal for this class of
models.

An illustrative example of a realistic point in parameter space has
spectrum as follows (all masses are in GeV):
$\chi_1^0 = 103$,
$\chi_1^\pm = 108$,
$h^0 = 128$,
$\chi_2^0 = 194$,
$\chi_3^0 = 205$,
$\chi_2^\pm = 220$,
$\tilde\ell \simeq 240$,
$H^0 = 266$,
$\chi_4^0 = 445$,
$\tilde{t} = 570, 700$,
$A^0 = 870$,
$\tilde{q} \simeq 1010$,
$\tilde{g} = 1030$,
tan$\beta = 9.1$.

\section{Supergravity and SUSY Radion Stabilization}
For the scenario above to work, it is crucial that the radion is stabilized
without SUSY breaking by a radion $F$ term.
Otherwise the leading contribution to visible scalar masses comes from
radion mediated SUSY breaking \cite{RMSB}; this also leads to viable
models with a phenomenology similar to gaugino mediated SUSY breaking
\cite{gMSB}.

We therefore analyze radion stabilization without SUSY breaking.
In the 4D effective theory, the radion $r$ is parameterized by a chiral
superfield with $\Re(T) \propto r$.
It is important to discuss the radion effective theory in the presence of
supergravity, since the \Kahler term for the radion field is proportional to
$T + T^\dagger$, which is trivial if supergravity effects are ignored.

We will consider theories in which the radion is stabilized in the 4D
effective theory, \ie $m_r \ll 1/r$.
In this case, we can write an effective 4D field theory of the radion
coupled to SUGRA:
\beq
\scr{L}_{\rm eff} = \myint d^4\th\, \phi^\dagger \phi f
+ \left( \myint d^2\th\, \phi^3 W + \hc \right),
\eeq
where $\phi = 1 + \th^2 F_\phi$ is the superconformal compensator
and $f = -3M_4^2 + \cdots$.
For the moment we consider a \Kahler function $f$ and superpotential
$W$ depending on a general set of 4D fields.
We will consider specific models below.

\subsection{Supersymmetric Vacua in Supergravity}
We first review the conditions to have a vacuum with
SUSY breaking and vanishing cosmological constant.
The supergravity potential is%
\footnote{Conventionally the supergravity potential is expressed in terms
of the \Kahler potential $K$, given by
$f = -3 M_4^2 e^{-K / 3 M_4^2}$.}
\beq[Vtrueus]
V = \frac{f_0^2}{f^2} \left[
(\tilde{f}^{-1})^a{}_b \left( W_a - \frac{3 f_a W}{f} \right)
\left( W^{\dagger b} - \frac{3 f^b W^\dagger}{f} \right)
+ \frac{9 |W|^2}{f} \right].
\eeq
where $f_0 = -3 M_4^2$ and
\beq
\tilde{f}^a{}_b = f^a{}_b - \frac{f^a f_b}{f}.
\eeq
The auxiliary fields are given by
\beq
F^\dagger_a = (\tilde{f}^{-1})_a{}^b \left[ W_b - \frac{3 f_b W}{f} \right],
\qquad
F_\phi = -\frac{1}{f} \left[ 3W + f^a F_a^\dagger \right].
\eeq
A vacuum has unbroken SUSY and vanishing cosmological
constant provided (assuming the matrix $\tilde{f}^a{}_b$ is nonsingular)
\beq
\frac{W}{f^3} = \hbox{\rm stationary},
\qquad
W = 0.
\eeq
This is satisfied provided that $W$ is stationary and $W = 0$ at the
minimum.
This is the well-known result that there is a one-to-one correspondence
between supersymmetric vacua in global SUSY and supergravity \cite{Weinberg}.

We now apply this to a theory where the only light field is the radion $T$.
The only possible subtlety is that the radion kinetic function
\beq
f  = -3 M_5^3 (T + T^\dagger)
\eeq
is unconventional.
However, $\tilde{f}^T{}_T = M_5^3 / (T + T^\dagger)$ is nonsingular for
nonzero $r$, and therefore the arguments above go through.
The conditions for radius stabilization with unbroken SUSY and vanishing
cosmological constant are therefore
\beq[Tvac]
\frac{\partial W}{\partial T} = 0,
\qquad
W = 0.
\eeq

\subsection{Bulk Super Yang--Mills}
A simple way of generating radion superpotentials is to consider a theory
with a super Yang-Mills (SYM) sector in the bulk.
Below the compactification scale, this becomes a $\scr{N} = 1$ SYM
theory (assuming an appropriate orbifold projection)
with a 4D gauge coupling that depends on $T$.
This theory becomes strong below the compactification scale and
generates a $T$-dependent dynamical superpotential \cite{LS1}
\beq
W_{\rm dyn} = a e^{-b T},
\eeq
where (for gauge group $SU(N)$)
\beq
a \sim \frac{1}{N^4 g_5^6},
\qquad
b = \frac{16\pi^2}{3 N g_5^2}.
\eeq
With a single SYM sector in the bulk, we cannot satisfy the conditions
\Eq{Tvac} for a SUSY vacuum.%
\footnote{\Ref{LS1} showed that this model with the addition of a constant
superpotential term and brane-localized SUSY breaking
stabilizes the radius with $\avg{F_T} \ne 0$.}
However, if there are two SYM sectors in the bulk, the dynamical
superpotential is
\beq
W_{\rm dyn} = a_1 e^{-b_1 T} + a_2 e^{-b_2 T},
\eeq
which has stationary points corresponding to SUSY vacua.
The condition $W = 0$ can be satisfied by adding a constant to the
superpotential
(which itself may arise dynamically from gaugino condensation).
The radion is stabilized at
\beq
\avg{T} = \frac{1}{b_1 - b_2} \ln\left( -\frac{a_1 b_1}{a_2 b_2} \right).
\eeq
$\avg{T}$ is real provided that $a_1$ and $a_2$ are real with opposite sign.

\section{Conclusions}
We have presented a simple and attractive mechanism for
obtaining realistic models of anomaly-mediated supersymmetry breaking.
The basis of the mechanism is that in four dimensions, $D$-type gauge
mediation gives flavor-blind, 2-loop contributions to the scalar masses
that are naturally of the same size as the 
contribution from anomaly mediation.
However in four dimensions it is generally
problematic to suppress direct contact terms between the hidden and visible
sectors which then give supersymmetry breaking contributions larger than
both these effects.
(See however \Ref{compdim}.)
We therefore embed the theory in a higher dimensional space,
with the visible and hidden sectors on different branes.
Now contact terms between the visible scalars and the hidden sector are
suppressed by higher-dimension locality, while contact terms between the
bulk gauge multiplets and the hidden sector are suppressed if the hidden
sector contains no singlets.
In this case, the visible scalars can obtain 2-loop $D$-type contribution
from `messenger' scalars localized on the hidden brane or in the bulk.
This mechanism preserves many of the attractive features of anomaly
mediation, in particular the predictions for the gaugino masses and
A-terms.

If the messengers are localized on the hidden brane, the $D$-type
contributions are the same size as the anomaly-mediated contributions
only for a special choice of compactification radius.
This is similar (and in fact closely related) to the mechanism of \Ref{GAMSB}.

We obtain a more robust and attractive model if the scalar messengers
are in the bulk.
In this case, the contributions to the soft scalar masses from $D$-type
gauge mediation and anomaly mediation are automatically the same size,
independent of the size of the extra dimension, and independent of whether
the theory is strongly or weakly coupled at the fundamental scale.
It is possible for the components of the supersymmetric gauge multiplet in
higher dimensions to play the role of the bulk messenger, so this mechanism
does not require the introduction of additional multiplets.

We have also shown that this framework
admits a solution to the $\mu$ problem based on
the next-to-minimal supersymmetric standard model.

In order for the effects we are considering to dominate over radion mediation,
we require that the brane spacing be stabilized in a supersymmetric way.
We construct a very simple model in which this naturally occurs.

We believe that this framework is the simplest possibility for
realistic models of anomaly mediated supersymmetry breaking.

\section*{Acknowledgments}

Z.C. was supported in part by the U.S. Department of Energy under Contract
DE-AC03-76SF00098, and in part by the NSF under grant PHY-00-98840. M.A.L.
was supported by the NSF under grant PHY-98-02551. Z.C. would like to 
thank Y. Nomura for a very useful discussion.



\begin{references}

\bibitem{RS0}
L.~Randall and R.~Sundrum,
``Out of this world supersymmetry breaking,''
Nucl.\ Phys.\ B {\bf 557}, 79 (1999)
[arXiv:hep-th/9810155].

\bibitem{compdim}
M.~A.~Luty and R.~Sundrum,
``Supersymmetry breaking and composite extra dimensions,''
arXiv:hep-th/0105137,
to be published in Phys.\ Rev.\ D;\\
M.~A.~Luty and R.~Sundrum,
``Anomaly mediated supersymmetry breaking in four dimensions,  naturally,''
arXiv:hep-th/0111231.

\bibitem{GLMR}
G.~F.~Giudice, M.~A.~Luty, H.~Murayama and R.~Rattazzi,
``Gaugino mass without singlets,''
JHEP {\bf 9812}, 027 (1998)
[arXiv:hep-ph/9810442].

\bibitem{GAMSB}
D.~E.~Kaplan and G.~D.~Kribs,
``Gaugino-assisted anomaly mediation,''
JHEP {\bf 0009}, 048 (2000)
[arXiv:hep-ph/0009195].

\bibitem{realAMSB} A.~Pomarol and R.~Rattazzi,
``Sparticle masses from the superconformal anomaly,''
JHEP {\bf 9905}, 013 (1999)
[arXiv:hep-ph/9903448]; \\
Z.~Chacko, M.~A.~Luty, I.~Maksymyk and E.~Ponton,
``Realistic anomaly-mediated supersymmetry breaking,''
JHEP {\bf 0004}, 001 (2000)
[arXiv:hep-ph/9905390]; \\
E.~Katz, Y.~Shadmi and Y.~Shirman,
``Heavy thresholds, slepton masses and the $\mu$ term in anomaly mediated
supersymmetry breaking,''
JHEP {\bf 9908}, 015 (1999)
[arXiv:hep-ph/9906296]; \\
I.~Jack and D.~R.~Jones,
``Fayet-Iliopoulos D-terms and anomaly mediated supersymmetry breaking,''
Phys.\ Lett.\ B {\bf 482}, 167 (2000)
[arXiv:hep-ph/0003081];\\
B.~C.~Allanach and A.~Dedes,
``R-parity violating anomaly mediated supersymmetry breaking,''
JHEP {\bf 0006}, 017 (2000)
[arXiv:hep-ph/0003222]; \\
Z.~Chacko, M.~A.~Luty, E.~Ponton, Y.~Shadmi and Y.~Shirman,
``The GUT scale and superpartner masses from anomaly mediated
supersymmetry breaking,''
Phys.\ Rev.\ D {\bf 64}, 055009 (2001)
[arXiv:hep-ph/0006047]; \\
N.~Arkani-Hamed, D.~E.~Kaplan, H.~Murayama and Y.~Nomura,
``Viable ultraviolet-insensitive supersymmetry breaking,''
JHEP {\bf 0102}, 041 (2001)
[arXiv:hep-ph/0012103].

\bibitem{experiment}
J.~L.~Feng, T.~Moroi, L.~Randall, M.~Strassler and S.~F.~Su,
``Discovering supersymmetry at the Tevatron in Wino LSP
scenarios,''
Phys.\ Rev.\ Lett.\ {\bf 83}, 1731 (1999)
[arXiv:hep-ph/9904250];\\
T.~Gherghetta, G.~F.~Giudice and J.~D.~Wells,
``Phenomenological consequences of supersymmetry with anomaly-induced
masses,''
Nucl.\ Phys.\ B {\bf 559}, 27 (1999)
[arXiv:hep-ph/9904378];\\
J.~L.~Feng and T.~Moroi,
``Supernatural supersymmetry: Phenomenological implications of
anomaly-mediated supersymmetry breaking,''
Phys.\ Rev.\ D {\bf 61}, 095004 (2000)
[arXiv:hep-ph/9907319]; \\
S.~Su,
``Higgs sector in anomaly-mediated supersymmetry breaking scenario,''
Nucl.\ Phys.\ B {\bf 573}, 87 (2000)
[arXiv:hep-ph/9910481]; \\
R.~Rattazzi, A.~Strumia and J.~D.~Wells,
``Phenomenology of deflected anomaly-mediation,''
Nucl.\ Phys.\ B {\bf 576}, 3 (2000)
[arXiv:hep-ph/9912390]; \\
D~.~K.~Ghosh, P.~Roy and S.~Roy,
``Linear collider signal of a Wino LSP in anomaly-mediated scenarios,''
JHEP {\bf 0008}, 031 (2000)
[arXiv:hep-ph/0004127]; \\
H.~Baer, J.~K.~Mizukoshi and X.~Tata,
``Reach of the CERN LHC for the minimal anomaly-mediated SUSY breaking
model,''
Phys.\ Lett.\ B {\bf 488},
367 (2000)
[arXiv:hep-ph/0007073]; \\
F.~E.~Paige and J.~Wells,
``Anomaly mediated SUSY breaking at the LHC,''
arXiv:hep-ph/0001249; \\
T.~Gherghetta, G.~F.~Giudice and J.~D.~Wells,
``Phenomenological consequences of supersymmetry with anomaly-induced
masses,''
Nucl.\ Phys.\ B {\bf 559}, 27 (1999)
[arXiv:hep-ph/9904378]; \\
D.~K.~Ghosh, A.~Kundu, P.~Roy and S.~Roy,
``Characteristic Wino signals in a linear collider from anomaly mediated
supersymmetry breaking,''
Phys.\ Rev.\ D {\bf 64}, 115001 (2001)
[arXiv:hep-ph/0104217].


\bibitem{MP}
E.~A.~Mirabelli and M.~E.~Peskin, ``Transmission of supersymmetry breaking
from a 4-dimensional boundary,'' Phys.\ Rev.\ D {\bf 58}, 065002 (1998)
[arXiv:hep-th/9712214].


\bibitem{AGW}
N.~Arkani-Hamed, L.~J.~Hall, D.~R.~Smith and N.~Weiner,
``Exponentially small supersymmetry breaking from extra dimensions,''
Phys.\ Rev.\ D {\bf 63}, 056003 (2001)
[arXiv:hep-ph/9911421]; \\
N.~Arkani-Hamed, T.~Gregoire and J.~Wacker,
``Higher dimensional supersymmetry in 4D superspace,''
arXiv:hep-th/0101233.

\bibitem{CLPU1}
Z.~Chacko, M.~A.~Luty and E.~Ponton,
``Massive higher-dimensional gauge fields as messengers of
supersymmetry  breaking,''
JHEP {\bf 0007}, 036 (2000)
[arXiv:hep-ph/9909248].

\bibitem{AHNSW} N.~Arkani-Hamed, L.~J.~Hall, Y.~Nomura,
D.~R.~Smith and N.~Weiner, ``Finite radiative electroweak symmetry
breaking from the bulk,'' Nucl.\ Phys.\ B {\bf 605}, 81 (2001)
[arXiv:hep-ph/0102090].

\bibitem{PT} E.~Poppitz and S.~P.~Trivedi,
``Some remarks on gauge-mediated supersymmetry breaking,''
Phys.\ Lett.\ B {\bf 401}, 38
(1997) [arXiv:hep-ph/9703246].


\bibitem{DineSeiberg}
M.~Dine and N.~Seiberg,
``Is The Superstring Weakly Coupled?,''
Phys.\ Lett.\ B {\bf 162}, 299 (1985).


\bibitem{RMSB}
Z.~Chacko and M.~A.~Luty,
``Radion mediated supersymmetry breaking,''
JHEP {\bf 0105}, 067 (2001)
[arXiv:hep-ph/0008103].

\bibitem{gMSB}
D.~E.~Kaplan, G.~D.~Kribs and M.~Schmaltz,
``Supersymmetry breaking through transparent extra dimensions,''
Phys.\ Rev.\ D {\bf 62}, 035010 (2000)
[arXiv:hep-ph/9911293];\\
Z.~Chacko, M.~A.~Luty, A.~E.~Nelson and E.~Ponton,
JHEP {\bf 0001}, 003 (2000)
[arXiv:hep-ph/9911323].

\bibitem{Weinberg}
S.~Weinberg,
``Does Gravitation Resolve The Ambiguity Among Supersymmetry Vacua?,''
Phys.\ Rev.\ Lett.\  {\bf 48}, 1776 (1982).

\bibitem{LS1}
M.~A.~Luty and R.~Sundrum,
``Radius stabilization and anomaly-mediated supersymmetry breaking,''
Phys.\ Rev.\ D {\bf 62}, 035008 (2000)
[arXiv:hep-th/9910202].

\end{references}
\end{document}